\documentclass[12pt]{iopart}

\usepackage{graphicx,iopams}

\begin{document}
\title[Comment on `Simulation of Bell states with incoherent thermal light']{Comment on `Simulation of Bell states with incoherent thermal light'}
\author{Jeffrey H Shapiro}
\address{Research Laboratory of Electronics,
Massachusetts Institute of Technology, Cambridge, MA 02139, USA}
\ead{jhs@mit.edu}

\begin{abstract}
Recently, Chen \em et al\/\rm.\ [New J. Phys. {\bf 13} (2011) 083018] presented experimental results, accompanied by quantum-mechanical analysis, showing that the quantum interference behavior of Bell states could be simulated in a modified Mach-Zehnder interferometer whose inputs are pseudothermal light beams obtained by passing laser light through a rotating ground-glass diffuser.  Their experiments and their theory presumed low-flux operation in which the simulated quantum interference is observed via photon-coincidence counting.  We first show that the Chen \em et al\/\rm.\ photon-coincidence counting experiments can be fully explained with semiclassical photodetection theory, in which light is taken to be a classical electromagnetic wave, and the discreteness of the electron charge leads to shot noise as the fundamental photodetection noise.  We then use semiclassical photodetection theory to show that the \em same\/\rm\ simulated quantum interference pattern can be observed in high-flux operation, when photocurrent cross-correlation is used instead of photon-coincidence counting.
\end{abstract}

\maketitle 
\section{Introduction}
The polarization Bell states of a pair of quantized spatiotemporal electromagnetic modes $A$ and $B$ are
\begin{equation}
|\psi^{\pm}\rangle \equiv \frac{|H\rangle_A|V\rangle_B \pm |V\rangle_A|H\rangle_B}{\sqrt{2}},
\end{equation}
and 
\begin{equation}
|\phi^{\pm}\rangle \equiv \frac{|H\rangle_A|H\rangle_B \pm |V\rangle_A|V\rangle_B}{\sqrt{2}},
\end{equation}
where $|H\rangle_K$ and $|V\rangle_K$, for $K=A,B$, denote single-photon states of horizontal and vertical polarization, respectively.  These states form a maximally-entangled basis for the two-qubit Hilbert space of single-photon states for modes $A$ and $B$.  As such, they are extremely important for applications such as quantum teleportation \cite{Teleport}, quantum superdense coding \cite{Superdense}, and quantum key distribution \cite{Ekert}, as well as their fundamental role in the Clauser-Horne-Shimony-Holt (CHSH) inequality \cite{CHSH}.  The standard approach to generating these states is to post-select the biphoton output from spontaneous parametric downconversion (SPDC) \cite{Kwiat}, and the standard approach to verify their entanglement behavior is via quantum-interference measurements \cite{Kwiat}.  Recently, Chen \em et al\/\rm.\ \cite{Chen} reported an experiment that mimicked the quantum-interference behavior seen with an SPDC entanglement source using two independent pseudothermal light beam obtained by passing laser light through a rotating ground-glass diffuser.  Their experiments were carried out in the low-flux regime using photon-coincidence counting, and they provided a quantum-mechanical explanation that ascribed their observations to two-photon interference, just as is the case for SPDC light.  

It has long been known that the semiclassical theory of photodetection---in which light is treated as a classical electromagnetic wave and the fundamental photodetection noise is the shot noise arising from the discreteness of the electron charge---produces quantitatively identical predictions to those obtained from quantum photodetection theory when the illumination is in a classical state, i.e., a coherent-state or a statistical mixture of such states.  See \cite{Shapiro} for a detailed review of this topic.  Except for any excess noise it may carry, laser light is coherent-state light.  Moreover, its propagation through ground-glass diffusers, free space, optical fibers, and beam splitters are all linear transformations, for which classical-state inputs yield classical-state outputs.  It follows that there must be an explanation for the experiments reported in \cite{Chen} that relies on semiclassical photodetection theory, i.e., one that only needs classical electromagnetic waves.   Our purpose in this paper is to present that explanation.  Furthermore, although we will begin with a treatment that applies to low-flux operation using photon-coincidence counting, our approach readily extends to high-flux operation using photocurrent cross-correlation.  

The remainder of the paper is organized as follows.  In Sec.~2 we describe the modified Mach-Zehnder interferometer---with photon-coincidence counting in the low-flux regime---that was employed in \cite{Chen}.  In Sec.~3 we introduce our classical-light model for this interferometer, and use it to derive the singles and coincidence rates as functions of the interferometer's differential time delay and its polarization-analysis angles.  Here we will show that our results explain the observations reported in \cite{Chen}.  Finally, in Sec.~4, we indicate how our theory can be extended to high-flux operation with photocurrent cross-correlation, and we provide some concluding remarks about the implications of our work.

\section{Modified Mach-Zehnder interferometer with pseudothermal Inputs}
The configuration for the experiment from \cite{Chen} is shown in figure~1.  A continuous-wave mode-locked Ti:sapphire laser operating at $\lambda = 780\,$nm wavelength with 78\,MHz pulse-repetition frequency and a $\tau_p\sim150\,$fs pulse duration illuminated an interference filter, to somewhat increase the pulse duration, followed by a rotating ground-glass diffuser, to render the light spatially incoherent.  The diameter $D \approx 5\,$mm output beam from the diffuser was divided by the 50-50 beam splitter BS1, with the resulting output beams propagating $d\approx\,200$mm (from the diffuser) to collection planes, one of which could be offset, longitudinally, by $\delta$.  Each collection plane contained the tip of a single-mode fiber, whose transverse coordinates, ${\boldsymbol \rho}_+$ and ${\boldsymbol \rho}_-$, satisfied $|{\boldsymbol \rho}_+-{\boldsymbol \rho}_-| \gg \ell_c$, where $\ell_c \approx 31\,\mu$m is the correlation length of the speckles cast in these planes.   The fibers routed the light they collected to polarizers P1 and P2, set for orthogonal polarizations that we shall take to be $x$ and $y$, respectively, before entering the 50-50 beam splitter BS2.  The outputs from BS2 then underwent polarization analysis, by analyzers A1 and A2 set for angles $\theta_A$ and $\theta_B$ with respect to $x$, prior to single-photon detection.  Computer processing completed the experiment by averaging the detector outputs over many pulses to obtain the singles rates $S_A$ and $S_B$ and the coincidence rate $C_{AB}$.  
\begin{figure}[h]
\begin{center}
\includegraphics[width=3.75in]{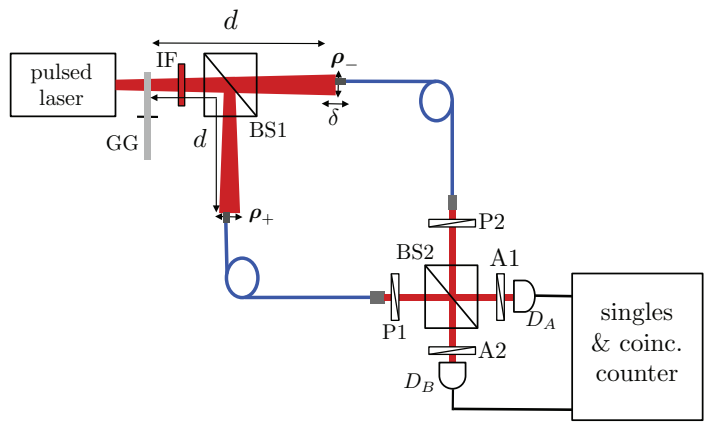}
\end{center}
\caption{Modified Mach-Zehnder interferometer with pseudothermal inputs, after \cite{Chen}. GG is a rotating ground-glass diffuser.  IF is an interference filter.  BS1 and BS2 are 50-50 beam splitters. $d$ is the nominal ground-glass to fiber-tip propagation distance, and $\delta$ is a variable offset in one arm.  P1 and P2 are polarizers that select orthogonal polarizations that we shall take to be $x$ and $y$, respectively.  A1 and A2 are polarization analyzers set for angles $\theta_A$ and $\theta_B$, respectively.  $D_A$ and $D_B$ are single-photon detectors}
\end{figure} 

Chen \em et al\/\rm.\ used quantum-mechanical analysis to show that $C_{AB}$ consisted of a background term plus a quantum interference term that is proportional to $\sin^2(\theta_A-\theta_B)$ when $\delta = 0$.  Their background term arises from what in a related prior experiment \cite{anticorrelation} they called ``self-intensity correlations.''  These correlations can be measured by summing the coincidence rates when a beam block is placed in front of the fiber tip at ${\boldsymbol \rho}_+$ and when a beam block is placed in front of the fiber tip at ${\boldsymbol \rho}_-$.  Subtracting the self-intensity correlations from the full coincidence rate obtained at $\delta = 0$ then yields, according to theory, a unity-visibility quantum interference pattern.  That background-subtracted quantum-interference pattern can then be used to obtain a violation of the CHSH inequality, in the same manner that is done \cite{Kwiat} for the polarization-entangled signal and idler obtained from an SPDC source.  The experimental data from \cite{Chen} bears out this theory:  the authors report $93.2\pm5.1$\% visibility in their background-subtracted quantum-interference pattern at $\delta = 0$.  

Chen \em et al\/\rm.\ do not attempt to explain their experiments with a classical-light model, i.e., with semiclassical photodetection theory.  In their previous work on a related anticorrelation experiment \cite{anticorrelation}, they claimed there was no classical-light explanation for their results.  We, however, have shown that such is not the case \cite{comment}.  In particular, we reported a straightforward classical-field analysis that reproduced the essential characteristics of the anticorrelation observed in \cite{anticorrelation}.  Our demonstration is especially important for the following reason.  Were Chen \em et al\/\rm.\ \cite{anticorrelation} correct in asserting that their anticorrelation measurement could not be explained in this manner it would present quantum optics with a major conundrum:  either laser light that has undergone linear transformation is \em not\/\rm\ in a coherent state or a random mixture of coherent states, or the quantum and semiclassical theories of photodetection \em can\/\rm\ make different quantitative predictions for the measurement statistics of classical-state light.  In the present paper, we shall extend our scalar-wave model from \cite{comment} to provide a classical-light explanation for the simulated Bell-state reported in \cite{Chen}.  

\section{Singles and coincidence rates}
The linear velocity of the ground-glass diffuser where it was illuminated in \cite{anticorrelation} was $\sim$0.8\,m/s, so that for the $\tau_p\sim 345$\,fs and $\tau_p\sim 541\,$fs post-IF pulse durations considered therein it is fair to say that the ground glass was completely stationary while a single laser pulse propagated through it.  We shall assume that to be the case for the experiments in \cite{Chen}.  The differential time delay $\delta t =\delta/c$ corresponding to the longitudinal variation $\delta$ was $\pm 4$\,ps in \cite{Chen} and we will assume that the photodetectors employed therein had the same $T\sim 1\,$ns coincidence gate as in \cite{anticorrelation}.  Hence, with $\tau_p < 1\,$ps, we have that $|\tau_p \pm \delta t| \ll T$.   Because the fibers are single mode, and because their transverse coordinates have been set to ensure that the light beams they collect are uncorrelated, we will assume (cf.\ \cite{comment}) that ${\bf E}_+(t)$ and ${\bf E}_-(t)$, the positive-frequency \em classical\/\rm\ fields emerging from polarizers P1 and P2, respectively, are given by 
\begin{equation}
{\bf E}_+(t) = v_+f(t+\delta t/2)e^{-i\omega_0t}{\bf i}_x,
\label{Eplus}
\end{equation}
and
\begin{equation}
{\bf E}_-(t) = v_-f(t-\delta t/2)e^{-i\omega_0t}{\bf i}_y.
\label{Eminus}
\end{equation}
Here: $v_+$ and $v_-$ are independent, identically distributed, zero-mean, isotropic, complex-valued Gaussian random variables with common mean-squared strength
\begin{equation}
\langle |v_+|^2\rangle = \langle |v_-|^2\rangle = N,
\end{equation}
representing constant-in-time speckle; 
\begin{equation}
f(t) \equiv \frac{e^{-t^2/\tau_p^2}}{(\pi \tau_p^2/2)^{1/4}},
\end{equation}
is a transform-limited Gaussian pulse normalized to satisfy
\begin{equation}
\int\!dt\,|f(t)|^2 = 1,
\end{equation}
with $\tau_p$ being the post-IF pulse duration;
and ${\bf i}_x$, ${\bf i}_y$ are orthogonal unit vectors along the $x$ and $y$ directions.  
Note that in (\ref{Eplus}) and (\ref{Eminus}) we have chosen $\sqrt{\mbox{photons/s}}$ units, so that the average energy in either field is $N\hbar\omega_0$, i.e., $N$ is their average photon number.   

Because Chen \em et al\/\rm.\ used photon-coincidence counting, their experiment was necessarily in the low-brightness regime wherein $N\ll 1$ prevails.  As a result, we can say that the singles rates (counts/gate) and the coincidence rate (coincidences/gate) obey \cite{Shapiro}
\begin{equation}
S_K = \eta \int_{-T/2}^{T/2}\!dt\,\langle |E_K(t)|^2\rangle ,\quad\mbox{for $K=A, B$},
\end{equation}
and
\begin{equation}
C_{AB} = \eta^2\int_{-T/2}^{T/2}\!dt\!\int_{-T/2}^{T/2}\!du\,\langle |E_A(t)|^2|E_B(u)|^2\rangle.
\end{equation}
In these expressions: $\eta$ is the photodetectors' quantum efficiency; the complex envelopes of the fields that illuminate detectors $D_A$ and $D_B$ are
\begin{equation}
E_A(t) = \frac{\cos(\theta_A)v_+f(t+\delta t/2) + \sin(\theta_A)v_-f(t-\delta t/2)}{\sqrt{2}},
\end{equation}
and
\begin{equation}
E_B(t) = \frac{\cos(\theta_B)v_+f(t+\delta t/2) - \sin(\theta_B)v_-f(t-\delta t/2)}{\sqrt{2}},
\end{equation}
where we have suppressed all propagation delays except for the interferometer's differential delay $\delta t$.  Using the statistics of $v_+$ and $v_-$, along with the disparities between $\tau_p$, $\delta t$, and $T$, it is easy to evaluate the singles rates and the coincidence rate.  

The statistical independence of the $v_+$ and $v_-$, and their mean-squared strengths, immediately gives us
\begin{eqnarray}
S_K &=& \frac{\eta }{2}\!\left(\langle |v_+|^2\rangle\cos^2(\theta_K)\int_{-T/2}^{T/2}\!dt\,|f(t+\delta t/2)|^2\right.  \nonumber \\[.05in]
&+&\left. \langle |v_-|^2\rangle \sin^2(\theta_K)\int_{-T/2}^{T/2}\!dt\,|f(t-\delta t/2)|^2\right) 
\\[.05in] &\approx& \frac{\eta N}{2}, \quad \mbox{for $K = A,B$},
\end{eqnarray}
where the approximation follows from $|\tau_p\pm \delta t/2| \ll T$.  Similarly, for the coincidence rate, the statistical independence of $v_+$ and $v_-$ leads to 
\begin{eqnarray}
\lefteqn{C_{AB} = \frac{\eta^2}{4}\!\left(c_A^2c_B^2 \langle|v_+|^4\rangle\int_{-T/2}^{T/2}\!dt\,|f(t_+)|^2\int_{-T/2}^{T/2}\!du\,|f(u_+)|^2\right.}\nonumber \\[.05in]
&+& c_A^2s_B^2\langle |v_+|^2\rangle \langle |v_-|^2\rangle \int_{-T/2}^{T/2}\!dt\,|f(t_+)|^2\int_{-T/2}^{T/2}\!du\,|f(u_-)|^2 \nonumber \\[.05in]
&+& c_B^2s_A^2\langle |v_+|^2\rangle \langle |v_-|^2\rangle\int_{-T/2}^{T/2}\!dt\,|f(t_-)|^2\int_{-T/2}^{T/2}\!du\,|f(u_+)|^2 \nonumber \\[.05in]
&-&2c_Ac_Bs_As_B\langle |v_+|^2\rangle \langle |v_-|^2\rangle\int_{-T/2}^{T/2}\!dt\!\int_{-T/2}^{T/2}\!du\,{\rm Re}[f^*(t_+)f^*(u_-)f(t_-)f(u_+)] \nonumber \\[.05in]
&+& \left.s_A^2s_B^2\langle |v_-|^4\rangle \int_{-T/2}^{T/2}\!dt\,|f(t_-)|^2\int_{-T/2}^{T/2}\!du\,|f(u_-)|^2\right),
\end{eqnarray}
where $c_K \equiv \cos(\theta_K)$ and $s_K \equiv \sin(\theta_K)$, for $K = A,B$, $w_\pm \equiv w\pm \delta t/2$, for $w = t, u$, and the first and last terms on the right are the self-intensity correlations.  Following \cite{Chen}, we shall suppress these self-intensity correlations---they can be found by first measuring the coincidence rate with ${\bf E}_+(t)$ blocked and then measuring the coincidence rate with ${\bf E}_-(t)$ blocked---by subtracting them from $C_{AB}$ and focus our attention on
\begin{eqnarray}
\lefteqn{\tilde{C}_{AB} = \frac{\eta^2N^2}{4}\!\left(c_A^2s_B^2\ \int_{-T/2}^{T/2}\!dt\,|f(t_+)|^2\int_{-T/2}^{T/2}\!du\,|f(u_-)|^2\right.} \nonumber \\[.05in]
&+& c_B^2s_A^2\int_{-T/2}^{T/2}\!dt\,|f(t_-)|^2\int_{-T/2}^{T/2}\!du\,|f(u_+)|^2 \nonumber \\[.05in]
&-&\left.2c_Ac_Bs_As_B\int_{-T/2}^{T/2}\!dt\!\int_{-T/2}^{T/2}\!du\,{\rm Re}[f^*(t_+)f^*(u_-)f(t_-)f(u_+)] \right),
\end{eqnarray}
where we have used the mean-squared values of $v_+$ and $v_-$.
Now, by exploiting $|\tau_p \pm \delta t/2| \ll T$ and some trigonometric identities, we can reduce the preceding expression to
\begin{equation}
\tilde{C}_{AB} \approx \frac{\eta^2 N^2}{4}\!\left[\sin^2(\theta_A - \theta_B) 
+ \frac{\sin(2\theta_A)\sin(2\theta_B)}{2}(1-e^{-\delta t^2/\tau_p^2})\right].
\label{Ctilde}
\end{equation}

At this point we are ready to compare our classical-light theory with the experimental results from \cite{Chen}.  There it is shown that $\theta_A = \theta_B = \pi/4$ leads to an anticorrelation dip of width $\sim$$\tau_p$ in $\tilde{C}_{AB}$ as $\delta t$ is scanned through zero.  Moreover, this anticorrelation dip has near-unity visibility, viz.,
\begin{equation}
\frac{\max(\tilde{C}_{AB}) - \min(\tilde{C}_{AB})}{\max(\tilde{C}_{AB}) + \min(\tilde{C}_{AB})} \approx 1.
\end{equation}
From (\ref{Ctilde}) we find
\begin{equation}
\tilde{C}_{AB} = \frac{\eta^2N^2}{8}(1-e^{-\delta t^2/\tau_p^2}),
\end{equation}
in agreement with those observations.  Likewise, Chen \em et al\/\rm.\ find that when $\theta_A = \pi/4$ and $\theta_B = 3\pi/4$ there is a near-unity visibility correlation peak of width $\sim$$\tau_p$ in $\tilde{C}_{AB}$ as $\delta t$ is scanned through zero.  From (\ref{Ctilde}) we get
\begin{equation}
\tilde{C}_{AB} = \frac{\eta^2N^2}{8}(1+e^{-\delta t^2/\tau_p^2}),
\end{equation}
for this case, as was found in the experiment.  Finally, at $\delta t = 0$, \cite{Chen} reports simulated Bell-state quantum-interference when $\theta_A-\theta_B$ is varied.  That characteristic is present in (\ref{Ctilde}), where we find
\begin{equation}
\tilde{C}_{AB} = \frac{\eta^2N^2}{4}\sin^2(\theta_A-\theta_B)
\end{equation}
when $\delta t = 0$.  Collectively, these matches between the Chen \em et al\/\rm.\ experimental results and the predictions of our classical-light theory are consistent with their experimental configuration's illuminating its detectors with classical-state light \em and\/\rm\ the quantitative agreement of the quantum and semiclassical theories of photodetection for such illumination.  
Because light is quantum mechanical, and photodetection is a quantum measurement, the quantum explanation of their work is of course the fundamental one, but we have just shown that it is \em not\/\rm\ necessary to invoke a quantum description of the light to explain their results.

\section{Discussion}
Our work in Sec.~3 presumed low-flux (single-photon) operation, specifically $N \ll 1$, in deriving the singles and coincidence rates for the figure~1 experiment.  Chen \em et al\/\rm.\ raised the question of what would happen if the illumination in this experiment was performed in the high-flux regime, wherein $N\gg 1$ prevails \cite{Chen}.  With our classical-light theory, it is easy to answer that question.  Suppose that ${\bf E}_{\pm}(t)$ are as given in Sec.~2 except that $N\gg 1$.  Also assume that the single-photon detectors in figure~1 are replaced by shot-noise limited PIN photodiodes, whose photocurrent outputs are $i_A(t)$ and $i_B(t)$, respectively, and that the singles and coincidence-rate measurements are replaced by average photocurrent and photocurrent cross-correlation measurements, $I^{(1)}_K \equiv \langle i_K(0)\rangle$, for $K= A,B$, and $I^{(2)}_{AB} \equiv \langle i_A(0)i_B(0)\rangle$.  We then have that \cite{Shapiro}
\begin{eqnarray}
\lefteqn{I^{(1)}_K = \frac{q\eta }{2}\!\left(\langle |v_+|^2\rangle\cos^2(\theta_K)\int\!d\tau\,|f(\tau+\delta t/2)|^2h(-\tau)\right. } \nonumber \\[.05in]
&+&\left. \langle |v_-|^2\rangle \sin^2(\theta_K)\int\!d\tau\,|f(\tau-\delta t/2)|^2h(-\tau)\right),
\end{eqnarray} 
where $q$ is the electron charge and $h(t)$ is the photodiodes' baseband impulse response, which is normalized to satisfy $\int\!dt\,h(t) = 1.$  To make the connection to the case of photon-coincidence counting, we shall ignore causality and take $h(t)$ to be
\begin{equation}
h(t) = \left\{\begin{array}{ll}
1/T, & \mbox{for $|t|\le T/2$}\\[.05in]
0, & \mbox{otherwise,}\end{array}\right.
\label{impulseresp}
\end{equation}
with $T\sim 1\,$ns, so that we again have $|\tau_p\pm \delta t/2| \ll T$.  We then can reduce our expression for the average photocurrents to the following simple form:
\begin{equation}
I^{(1)}_K \approx \frac{q\eta N}{2T}, \quad \mbox{for $K = A,B$}.
\end{equation}

For the photocurrent cross-correlation we have 
\begin{eqnarray}
\lefteqn{I^{(2)}_{AB} = \frac{q^2\eta^2}{4T^2}\!\left(c_A^2c_B^2 \langle|v_+|^4\rangle\int_{-T/2}^{T/2}\!dt\,|f(t_+)|^2\int_{-T/2}^{T/2}\!du\,|f(u_+)|^2\right.}\nonumber \\[.05in]
&+& c_A^2s_B^2\langle |v_+|^2\rangle \langle |v_-|^2\rangle \int_{-T/2}^{T/2}\!dt\,|f(t_+)|^2\int_{-T/2}^{T/2}\!du\,|f(u_-)|^2 \nonumber \\[.05in]
&+& c_B^2s_A^2\langle |v_+|^2\rangle \langle |v_-|^2\rangle\int_{-T/2}^{T/2}\!dt\,|f(t_-)|^2\int_{-T/2}^{T/2}\!du\,|f(u_+)|^2 \nonumber \\[.05in]
&-&2c_Ac_Bs_As_B\langle |v_+|^2\rangle \langle |v_-|^2\rangle\int_{-T/2}^{T/2}\!dt\!\int_{-T/2}^{T/2}\!du\,{\rm Re}[f^*(t_+)f^*(u_-)f(t_-)f(u_+)] \nonumber \\[.05in]
&+& \left.s_A^2s_B^2\langle |v_-|^4\rangle \int_{-T/2}^{T/2}\!dt\,|f(t_-)|^2\int_{-T/2}^{T/2}\!du\,|f(u_-)|^2\right),
\end{eqnarray}
where we have used (\ref{impulseresp}) for $h(t)$, and the first and last terms on the right come from self-intensity correlations.  Subtracting the self-intensity correlations from $I^{(2)}_{AB}$, and using $\tilde{I}^{(2)}_{AB}$ to denote the result, we arrive at
\begin{eqnarray}
\lefteqn{\tilde{I}^{(2)}_{AB} = \frac{q^2\eta^2N^2}{4T^2}\!\left(c_A^2s_B^2\ \int_{-T/2}^{T/2}\!dt\,|f(t_+)|^2\int_{-T/2}^{T/2}\!du\,|f(u_-)|^2\right.} \nonumber \\[.05in]
&+& c_B^2s_A^2\int_{-T/2}^{T/2}\!dt\,|f(t_-)|^2\int_{-T/2}^{T/2}\!du\,|f(u_+)|^2 \nonumber \\[.05in]
&-&\left.2c_Ac_Bs_As_B\int_{-T/2}^{T/2}\!dt\!\int_{-T/2}^{T/2}\!du\,{\rm Re}[f^*(t_+)f^*(u_-)f(t_-)f(u_+)] \right) \\[.05in]
&\approx& \frac{q^2\eta^2 N^2}{4T^2}\!\left[\sin^2(\theta_A - \theta_B) 
+ \frac{\sin(2\theta_A)\sin(2\theta_B)}{2}(1-e^{-\delta t^2/\tau_p^2})\right].
\end{eqnarray}
Thus the \em same\/\rm\ simulated Bell-state behavior seen in photon-coincidence counting in low-flux operation will be present in the photocurrent cross-correlation in high-flux operation.

Let us conclude with a brief meta-lesson that can be gleaned from what we have done.  Chen \em et al\/\rm.\ \cite{Chen} use quantum analysis to show that photon-coincidence counting in the figure~1 configuration---after subtraction of self-intensity correlations---mimics Bell-state quantum interference.  Because classical-state light is employed in their experiments, their observations have an equivalent quantitative explanation from semiclassical photodetection theory.  Moreover, as we have just seen, that same mimicking of Bell-state quantum interference will be present in photocurrent cross-correlations, taken in high-flux operation, after self-intensity correlations are subtracted.  In the high-flux regime, it is clear that semiclassical theory ascribes the interference pattern to intensity-fluctuation correlations between $|E_A(t)|^2$ and $|E_B(t)|^2$.  Inasmuch as the experimental results from \cite{Chen} \em cannot\/\rm\ distinguish between the quantum and semiclassical theories, it is disingenuous to claim that the figure~1 configuration generates Bell states, as opposed to simulating the quantum-interference pattern produced by detection of a Bell state.  After all, Bell states are entangled states, and entangled states are nonclassical.  Hence their photodetection statistics \em cannot\/\rm\ be fully and properly quantified by semiclassical photodetection theory.  In this regard we note that post-selection, which is used to identify Bell-state photon pairs produced by SPDC, is very different from subtracting the self-intensity correlations in the figure~1 experiment to see Bell-state quantum interference.  This is because post-selection is just selecting the occurrence of a photon coincidence, i.e., it can be performed on a pulse-by-pulse basis, whereas subtracting the self-intensity correlations in figure~1 requires collecting averages of coincidences with both beams present at BS2 and with only one beam or the other present on that beam splitter.  

\ack
This work was sponsored by the DARPA Information in a Photon Program under U.S. Army Research Office Grant No.\ W911NF-10-1-0404.\\


\section*{References}

\begin{thebibliography}{99}

\bibitem{Teleport}Bennett C H, Brassard G, Cr\'{e}peau C, Jozsa R, Peres A and Wootters W K 1993 \PRL {\bf 70} 1895
\bibitem{Superdense}Bennett C and Wiesner S J 1992 \PRL {\bf 69} 2881  
\bibitem{Ekert}Ekert A K 1991 \PRL {\bf 88} 057902
\bibitem{CHSH}Clauser J F, Horne M, Shimony, A and Holt R A 1969 \PRL {\bf 23} 880
\bibitem{Kwiat}Kwiat P G, Mattle K, Weinfurter H, Zeilinger A, Sergienko A V and Shih Y 1995 \PRL 4337
\bibitem{Chen}Chen H, Peng T, Karmakar S and Shih Y 2011 {\it New J. Phys.} {\bf 13} 083018
\bibitem{Shapiro}Shapiro J H 2009 {\it IEEE J. Sel. Top. Quantum Electron.} {\bf 15} 1547
\bibitem{anticorrelation}Chen H, Peng T, Karmakar, S, Xie, Z and Shih Y 2011 {\it Phys. Rev. A}  {\bf 84} 033835
\bibitem{comment}Shapiro J H 2011 arXiv:1110.5691 [quant-ph]

\end{thebibliography}
\end{document}